\documentstyle[12pt]{article}
\huge
\renewcommand{\theequation}{\arabic{section}.\arabic{equation}} 
\def\setzero{\setcounter{equation}{0}}

\newcounter{eqalph}
\def\bph{\setcounter{eqalph}{\value{equation}}
   \addtocounter{eqalph}{1}
   \setcounter{equation}{0}
   \renewcommand{\theequation}{\arabic{section}.\arabic{eqalph}\alph{equation}}}
\def\eph{\setcounter{equation}{\value{eqalph}}
   \renewcommand{\theequation}{\arabic{section}.\arabic{equation}}
\par\noindent}

\begin{document}

\baselineskip 18pt

\def \sech{{\rm sech}}
\def \tanh{{\rm tanh}}
\def \cn{{\rm cn}}
\def \sn{{\rm sn}}
\def\bm#1{\mbox{\boldmath $#1$}}
\newfont{\bg}{cmr10 scaled\magstep4}
\newcommand{\bzr}{\smash{\hbox{\bg 0}}}
\newcommand{\bzl}{%
   \smash{\lower1.7ex\hbox{\bg 0}}}
\title{Matrix Models of Two-Dimensional Gravity and Discrete Toda Theory} 
\date{\today}
\author{ Masato {\sc Hisakado} \footnotemark[1]  \hspace{2 mm} and Miki {\sc Wadati}
\footnotemark[2]
\\
\bigskip
\\ \footnotemark[1]
{\small\it Graduate School  of Mathematical Sciences,}\\
{\small\it University of Tokyo,}\\
{\small\it 3-8-1 Komaba, Megro-ku, Tokyo, 153, Japan}
\vspace{10 mm}
\\ \footnotemark[2]
{\small\it Department of Physics, Faculty of Science,}\\
{\small\it University of Tokyo,}\\
{\small\it 7-3-1 Hongo, Bunkyo-ku, Tokyo, 113, Japan}}

\maketitle

\vspace{20 mm}

Abstract

Recursion relations for orthogonal polynominals, arising in the study 
of  the one-matrix model of two-dimensional gravity, are shown to be 
equvalent to the equations of the Toda-chain  hierarchy 
supplemented by  additional Virasoro constraints.
This is  the case without  the double scaling limit.
 A discrete time variable to the matrix model is introduced.
 The discrete time dependent partition functions are given 
by $\tau$ functions which satisfy the discrete Toda molecule equation.
Further  the relations between the matrix model and 
the discrete time Toda theory are discussed.

\vfill

\newpage

\section{Introduction}
Matrix models have 
 attracted  a lot of attention  as possible  models for 
nonperturbative  string theory and  two-dimensional gravity.
In particuler the matrix model of two-dimensional gravity can be solved exactly using the 
orthogonal polynominal theory.\cite{b}
One of the most intriguing features is the  connection
with the theory of integrable systems. It has been shown  that the partition 
function of a matrix model is the  $\tau$-function.
Thus  we see that an important  object
in the theory of integrable equations appears naturally
 from the matrix model.

 Discrete-time integrable system is  currently a 
focus of an intensive  study.
It is interesting  to note that most of the well-known features of 
continuous-time integrable systems  such as Lax-pair, B\"{a}cklund 
transformation, Painlev\'{e} property, etc., carry over  to the case 
of discrete-time integrable systems.
The Toda molecule equation with discrete time (hereafter, the discrete Toda molecule equation)  can be utilized for the numerical computation of matrix 
eigenvalues.
Hirota  {\it et al}. have  suggested similarities between their method 
and the $LR$ factorization method.\cite{h}

Let us introduce the discrete Toda molecule equation.
Let  $I_{0},\cdots,I_{N-1}$ and $V_{0},\cdots,V_{N-2}$ be   dynamical 
variables, and $l$ be discrete ``time'' variable.
The discrete-time Toda molecule equation 
is given  by,\cite{h}
\begin{equation}
I_{k}(l+1)-I_{k}(l)=V_{k}(l)-V_{k-1}(l+1),
\label{1}
\end{equation}
\begin{equation}
I_{k}(l+1)V_{k}(l+1)=I_{k+1}(l)V_{k}(l).
\label{2}
\end{equation}
 The boundary condition $V_{-1}=V_{N-1}=0$ is imposed 
for any time $l$.
By a transformation, 
\begin{eqnarray}
I_{k}(l)&=&\frac{\tau_{k}(l+1)\tau_{k-1}(l)}{\tau_{k}(l)\tau_{k-1}(l+1)},
\nonumber \\
V_{k}(l)&=&\frac{\tau_{k+1}(l)\tau_{k-1}(l+1)}{\tau_{k}(l)\tau_{k}(l+1)},
\label{3}
\end{eqnarray}
the equations of motion, (\ref{1}) and (\ref{2}), are cast 
into the following bilinear equation,
\begin{equation}
\tau_{k}(l)^{2}-\tau_{k}(l-1)\tau_{k}(l+1)+\tau_{k+1}(l-1)\tau_{k-1}
(l+1)=0.
\label{4}
\end{equation}
One easily  see  that under (\ref{3}), (\ref{2}) holds   identically 
 and (\ref{1}) reduces to (\ref{4}).

The outline of this letter is as follows. 
In  section 2 we first consider a theory of discrete-time dependent
orthonormal polynominals (DTOP).
Then we  see that  a case of DTOP can be formulated 
by the discrete Toda theory.
The Lax-pair of the discrete Toda molecule equation is derived using the 
properties of the orthonormal polynominals.
In  section 3 we show our $\tau$  function in  section 2 is the partition function 
of the scaler product model.
This  model can be obtained in the limit of the Kontsevich integral. 
In  section 4 we extend Hirota's discrete Toda theory 
and  apply the discrete Toda theory to DTOP.
In section 5 
we show a way from  DTOP to  
string  equation.
In section 6 
we consider about  continuous and discrete Toda theories.
The last section is devoted to concluding remarks.
\setzero

\section{Discrete Toda Theory and Orthogonal Polynominals}

Let us begin by associating  a time variable $l$ to the orthogonal
 polynominal theory.\cite{s},\cite{g}
The distribution function $\rho(\lambda;l)$ or the weight function
 $w(\lambda;l)={\rm d}\rho(\lambda;l)/{\rm d}\lambda$ 
uniquely defines the orthonormal polynominal $\varphi_{n}(\lambda;l)$.
 According to Schmidt's diagonalization method, we have 
\begin{eqnarray}
\varphi_{n}(\lambda;l)=\frac{1}{\sqrt{\tau_{n}\tau_{n-1}}}
\left|
\begin{array}{cccc}
s_{0}&s_{1}&\cdots&s_{n} \\
s_{1}&s_{2}&\cdots&s_{n+1}\\
\cdots&\cdots&\cdots&\cdots\\
s_{n-1}&s_{n}&\cdots&s_{2n-1}\\
1&\lambda&\cdots&\lambda^{n}
\end{array}
\right|,
\label{2.1}
\end{eqnarray}
where $s_{n}$'s are the moments defined by $s_{n}(l)=\int\lambda^{n}
{\rm d}\rho(\lambda;l)$  and the tau function $\tau_{n}$ is defined by
\begin{eqnarray}
\tau_{n}(l)=
\left|
\begin{array}{cccc}
s_{0}&s_{1}&\cdots&s_{n}\\
s_{1}&s_{2}&\cdots&s_{n+1}\\
\cdots&\cdots&\cdots&\cdots\\
s_{n}&s_{n+1}&\cdots&s_{2n}
\end{array}
\right|.
\label{tau}
\end{eqnarray}
Here we consider the case $w(\lambda;l+1)=\lambda w(\lambda;l)$.
Remark that the determinant in (\ref{2.1})  is  a Hankel-Hadamard-type determinant.
$\varphi_{n}(\lambda;l)$'s   satisfy the orthonormal relations,
\begin{equation}
\int\varphi_{n}(\lambda;l)\varphi_{m}^{*}(\lambda;l){\rm d}\rho(\lambda;l)=\delta_{mn},\;\;
l=1,2,\cdots.
\label{sp}
\end{equation}
In matrix models based on the  hermitian matrices the integral 
in (\ref{sp}) is along the real axis,  and therefore  $\varphi_{m}^{*}=\varphi_{m}$ are  assumed  hereafter.
These  are basic results  of the orthogonal 
polynominal theory, except that we have introduced  a discrete-time 
 (``zero-time'') variable $l$. This tau function is precisely 
the same tau function as in Hirota-Sato's soliton theory.
The tau function $\tau_{n}(l)$ satisfies the discrete Toda molecule  equation (\ref{4}),
 which is known to be  a completely
 integrable system. 

We derive the Lax-pair of this system using  the
orthogonal polynominals.
First, we know from the orthogonal polynominal theory  that 
the three-term relationship holds:
\begin{equation}
a_{n+1}(l)\varphi_{n+1}(\lambda;l)+b_{n}(l)\varphi_{n}(\lambda;l)+a_{n}(l)
\varphi_{n-1}(\lambda;l)=\lambda\varphi_{n}(\lambda;l),
\label{l1}
\end{equation}
where $a_{n}$'s and $b_{n}$'s are given by
\begin{eqnarray}
a_{n}(l)&=&\int\lambda\varphi_{n}(\lambda;l)\varphi_{n-1}(\lambda;l){\rm d}\rho(\lambda;l),
\nonumber \\
b_{n}(l)&=&\int\lambda\varphi_{n}^{2}(\lambda;l){\rm d}\rho(\lambda;l).
\end{eqnarray}
It is a half (space part) of the Lax-pair.
By using (\ref{sp}) and the condition 
$w(\lambda;l+1)=\lambda w(\lambda;l)$ the other half (time part) is obtained,
\begin{equation}
\alpha_{n}(l)\varphi_{n}(\lambda;l)=\alpha_{n}(l+1)\varphi_{n}(\lambda;l+1)
+V_{n-1}(l)\alpha_{n-1}(l+1)\varphi_{n-1}(\lambda;l+1),
\label{l2}
\end{equation}
where 
\begin{equation}
\frac{\alpha_{n+1}(l)}{\alpha_{n}(l)}=a_{n+1}(l),
\end{equation}
\begin{equation}
a_{n}^{2}(l)=I_{n-1}(l)V_{n-1}(l),
\label{a}
\end{equation}
\begin{equation}
b_{n}(l)=I_{n}(l)+V_{n-1}(l).
\label{b}
\end{equation}
As a compatibility condition of (\ref{l1}) and (\ref{l2}), we obtain 
the equations of motion 
 (\ref{1}) and (\ref{2}) 
for field variables $I_{n}$ and $V_{n}$. 

{}From (\ref{3}) and (\ref{a}) we have 
\begin{equation} 
a_{n+1}^{2}(l)=\frac{\tau_{n+1}(l)\tau_{n-1}(l)}{\tau_{n}^{2}(l)}.
\label{ta}
\end{equation}
Note that (\ref{ta}) is the same for  the continuous-time Toda molecule  equation.
A useful  formula which expresss  $\tau_{n}$ in termes of $a_{k}$'s 
is derived by inverting   (\ref{ta})
,
\begin{equation}
\tau_{n}=s_{0}^{n+1}(a_{1}^{2})^{n}(a_{2}^{2})^{n-1}\cdots(a_{n}^{2})^{1}.
\label{bt}
\end{equation}
This formula is called  a Barners-type formula.\cite{s}

\setzero
\section{Matrix Model and Discrete Toda Theory}

For $p\times p$ hermitian matrices $X$ and $K$, let us consider  
the Kontsevich integral,\cite{w},\cite{k},\cite{m2}
\begin{equation}
{\cal F}_{V,p}\{K\}=\int_{p\times p}{\rm d}Xe^{-{\rm tr}V(X)+{\rm tr}KX}.
\label{ci}
\end{equation}
The eigenvalues of matrices $X$ and $K$ are respectively denoted as $x_{\gamma}$ and $k_{\delta}$.
Then the Kontsevich integral (\ref{ci}) is  calculated as \cite{m}
\begin{equation}
{\cal F}_{V,p}\{K\}=(2\pi)^{p(p-1)/2}\frac{{\rm det}_{\gamma\delta}\hat
{\phi}_{\gamma}(k_{\delta})}{\Delta(k)},
\end{equation}
where 
\begin{equation}
\hat{\phi}_{\gamma}(k)=\int dxx^{\gamma-1}e^{-V(x)+kx},\;\;\gamma=1,2,\cdots,p
\end{equation}
and $\Delta(k)$ is the ``van der Monde determinant'', $\Delta(k)=
{\rm det}_{ij}k_{i}^{j-1}=\Pi_{i>j}
^{N}(k_{i}-k_{j})$.
Note  that if the ``zero-time'' $l$ is introduced, then
\begin{equation}
{\cal F}_{V,p}\{l|K\}={\cal F}_{V(X)-l{\rm log}X, p}\{K\}=(2\pi)^{p(p-1)/2}\frac{{\rm det}_{\gamma\delta}\hat
{\phi}_{\gamma}(k_{\delta},l)}{\Delta(k)},
\end{equation}
where
\begin{equation}
\hat{\phi}_{\gamma}(k,l)=\int dxx^{l+\gamma-1}e^{-V(x)+kx}.
\end{equation}

We introduce   one matrix model  of the form:
\begin{equation}
Z_{p}=c_{p}\int_{p\times p}dX  e^{-{\rm Tr}V(X)},
\end{equation}
where $c_{p}$ is a constant.

Taking $V=V(X)-l\log X$,  we get, 
\begin{eqnarray}
Z_{p}(l)&=&c_{p}\int_{p\times p}dXe^{-{\rm Tr}V(X)}=\lim_{K\rightarrow 0}
{\cal F}_{V,p}\{l|K\}
=\lim_{\{k_{j}\}\rightarrow0}
\frac{{\rm Det}_{ij}\hat{\phi}_{i}^{\{V\}}(k_{j},l)}{\Delta(k)}
\nonumber \\
&=&{\rm Det}_{ij}\frac{\partial^{j-1}\hat{\phi}_{i}^{\{V\}}(k_{j},l)}
{\partial k^{j-1}}|_{0}={\rm Det}_{ij}(\frac{\partial}{\partial k})^{i+j-2}
\hat{\Phi}^{\{V\}}|_{0}.
\label{km}
\end{eqnarray}
 Note that $\hat{\Phi}^{\{V\}}(l)=\int dxx^{l}e^{-V(x)+kx}$.
The formula (\ref{km}) is the same as (\ref{tau}) in the case
\begin{equation}
\frac{{\rm d}\rho(\lambda;l)}{{\rm d}\lambda}=\lambda^{l}\exp(-V(\lambda)).
\end{equation}
Then,  one  find that the partition function of the one matrix  model is nothing but the  tau function.
In the special case $l=0$ the one matrix  is the normal 
one-matrix model. 
\setzero
\section{Extended Discrete Toda Theory}
We choose  the orthogonal  polynomimals as follows:
\begin{equation}
\Phi_{n}(\lambda;l)\equiv e^{\phi_{n}(l)/2}\varphi_{n}(\lambda;l)=\alpha_{n}(l)\varphi_{n}(\lambda;l),
\end{equation}
where
\begin{equation}
a_{n}(l)=e^{\phi_{n}(l)/2-\phi_{n-1}(l)/2}.
\end{equation}
$\Phi_{n}(\lambda;l)$ satisfy a trivial scalar product,
\begin{equation}
\int\Phi_{n}(\lambda;l)\Phi_{m}(\lambda;l){\rm d}\rho(\lambda;l)=e^{\phi_{n}(l)}\delta_{nm}.
\end{equation}
Recalling  a Barners-type formula (\ref{bt}), we see that the partition function of the one matrix model 
is given by
\begin{equation}
Z_{p}(l)=\prod_{i=0}^{p-1}e^{\phi_{i}(l)}.
\end{equation}

We can rewrite representations (\ref{l1}) and (\ref{l2}) as 
\begin{equation}
\lambda\Phi_{n}(\lambda;l)=\Phi_{n+1}(\lambda;l)+\{I_{n}(l)+V_{n-1}(l)\}\Phi_{n}(\lambda;l)
+V_{n-1}(l)I_{n-1}(l)\Phi_{n-1}(\lambda;l),
\label{L1}
\end{equation}
\begin{equation}
\Phi_{n}(\lambda;l)=\Phi_{n}(\lambda;l+1)+V_{n-1}(l)\Phi_{n-1}(\lambda;l+1),
\label{L2}
\end{equation}
for $n=0,1,2,\cdots, N-1$.
In fact we have to consider $N$ in the limit $N\rightarrow \infty$ hereafter.
However for convinience'sake, we use $N$.
{}From (\ref{L1}) and (\ref{L2}) we have
\begin{equation}
\lambda\Phi_{n}(\lambda;l+1)=I_{n}(l)\Phi_{n}(\lambda;l)+\Phi_{n+1}(\lambda;l).
\label{L3}
\end{equation}
Note that (\ref{L2}) and (\ref{L3}) reduces to  Hirota's operators \cite{h} if we set $\lambda=1$.
Let  $|l\rangle$ be an $N$-dimensional vector:
\begin{equation}
|l\rangle=[\Phi_{0}(\lambda;l),\Phi_{1}(\lambda;l),\cdots,\Phi_{N-1}(\lambda;l)]^{T},
\end{equation}
where the superscript $T$ denotes the transposition of  matrix, and $L(l)$ and $R(l)$ be 
$N\times N$ matrices:
\begin{eqnarray}
L(l)=
\left(
\begin{array}{ccccc}
1& & & &\bzr\\
V_{0}(l)&1& & & \\
 &V_{1}(l)& 1& & \\
 &      &\ddots&\ddots& \\
\bzl&  &  &V_{N-2}(l)&1
\end{array}
\right),
\label{k1}
\end{eqnarray}
\begin{eqnarray}
R(l)=
\left(
\begin{array}{ccccc}
I_{0}(l)&1& &  & \bzr\\
 &I_{1}(l)&1&  &  \\
 &       &I_{2}(l)&\ddots & \\
  &      &  & \ddots&1\\
\bzl&  & &  &I_{N-1}(l)
\end{array}
\right).
\label{k2}
\end{eqnarray}
In terms of the matrices $L(l)$ and $R(l)$, (\ref{L2}) and (\ref{L3}) are expressed as 
\begin{equation}
|l\rangle=L(l)|l+1\rangle,
\label{b1}
\end{equation}
\begin{equation}
\lambda|l+1\rangle=R(l)|l\rangle.
\label{b2}
\end{equation}
Let us introduc a matrix  $A(l)$  by
\begin{equation}
A(l)=L(l)R(l).
\label{b3}
\end{equation}

{}From (\ref{b1})-(\ref{b3}), we  have 
\begin{equation}
\lambda^{k}|l\rangle=A(l)^{k}|l\rangle.
\label{l6}
\end{equation}
We define new operators 
\begin{equation}
L^{(k)}(l)=L(l)L(l+1)\cdots L(l+k-1),
\label{lk}
\end{equation}
\begin{equation}
R^{(k)}(l)=R(l+k-1)R(l+k-2)\cdots R(l).
\label{rk}
\end{equation}
{}From (\ref{b1})-(\ref{b3}), we also have
\bph
\begin{equation}
L^{(k)}(l)|l+k\rangle=|l\rangle,
\end{equation}
\begin{equation}
R^{(k)}(l)|l\rangle=\lambda^{k}|l+k\rangle,
\end{equation}
\begin{equation}
(L^{(k)}(l))^{-1}L^{(k)}(l)=E,\;\;
\end{equation}
\eph
where $E$ is an identity operator.

The compatibility condition of (\ref{l6}) and (4.17b) yields a 
matrix equation:
\begin{equation}
A(l+k)^{k}=(L^{(k)}(l))^{-1}A(l)^{k}L^{(k)}(l).
\label{dth}
\end{equation}
In the paticular case $k=1$, (\ref{dth}) becomes
\begin{equation}
R(l)L(l)=L(l+1)R(l+1),
\end{equation}
which gives the discrete Toda molecule equation.

We may use the bra-ket expression for scalar product,
\begin{equation}
\langle \Phi_{n}(\lambda;l)|\Phi_{m}(\lambda;k)\rangle
=\int\Phi_{n}(\lambda;l)\Phi_{m}(\lambda;k){\rm d}\rho(0).
\label{ipp}
\end{equation}
Remark that the distribution function  is $\rho(0)\equiv\rho(0;l)$ in all $l$.
Then, the following relations hold, 
\begin{eqnarray}
e^{\phi_{n}(l)}&=&\int\Phi_{n}(\lambda;l)^{2}{\rm d}\rho(\lambda;l)=\langle\Phi_{n}(\lambda;l)
|\lambda^{l}|\Phi_{n}(\lambda;l)\rangle 
\nonumber \\
&=&
\langle \Phi_{n}(\lambda;l)|A^{l}|\Phi_{n}(\lambda;l)\rangle.
\end{eqnarray}
\setzero
\section{String equation}
In this section we consider the one matrix  model 
\begin{equation}
Z_{p}(l,t)\equiv c_{p}\int_{p\times p}dH ({\rm det}H)^{l}e^{-\sum_{k=0}
^{\infty}t_{k}{\rm Tr}H^{k}},
\end{equation}
where $H$ is a $p\times p$ hermitian matrix.
 For this model the string equation is as follows:\cite{m1},\cite{m2},\cite{m}
\begin{equation}
L_{q}(l)Z_{p}(l)=0,\;\;q\geq-1,
\label{sde1}
\end{equation}
with
\begin{eqnarray}
L_{q}(l)&\equiv&\sum_{k=0}^{\infty}kt_{k}\frac{\partial}{\partial t_{k+q}}
+\sum_{k=0}^{q}\frac{\partial^{2}}{\partial t_{k}\partial t_{q-k}}
-l\frac{\partial}{\partial t_{q}}\;\;\;\;q\geq0,
\nonumber \\
L_{-1}(l)&\equiv&\sum_{k=0}^{\infty}kt_{k}\frac{\partial}{\partial t_{k-1}}
-l\frac{\partial}{\partial t_{-1}}.
\label{sde}
\end{eqnarray}
Note that $\partial/\partial t_{-1}$ and $\partial/\partial t_{0}$ 
are defined:
\begin{equation}
\frac{\partial}{\partial t_{-1}}e^{\phi_{n}(l)}
=\langle\Phi_{n}(\lambda;l)|\lambda^{l-1}|\Phi_{n}(\lambda;l)\rangle,
\end{equation}
\begin{equation}
\frac{\partial}{\partial t_{0}}e^{\phi_{n}(l)}
=\langle\Phi_{n}(\lambda;l)|A^{l}|\Phi_{n}(\lambda;l)\rangle=e^{\phi_{n}(l)}.
\label{d0}
\end{equation}

Using  (\ref{d0}), we have 
\begin{equation}
\frac{\partial}{\partial t_{0}}Z_{p}(l)=pZ_{p}(l).
\end{equation}
It is easy to see that $L_{n}$'s form a closed Virasoro algebra:
\begin{equation}
[L_{n}(l),L_{m}(l)]=(n-m)L_{n+m}(l),\;\;n,m\geq0.
\end{equation}
They are identified with the Virasoro constraints for a normal one-matrix model
 without $l$.

Let us discuss the  string  equation from the theory orthonormal of polynominals.
 We can show  that the weight function 
 is satisfied  if
\begin{equation}
\frac{{\rm d}\rho(\lambda;l)}{{\rm d}\lambda}
=\lambda^{l}\exp (-\sum_{k=0}^{\infty}t_{k}\lambda^{k}).
\end{equation}

First we have, from the orthonormality,
\begin{eqnarray}
0&=&\int\frac{\partial \varphi_{n}(\lambda;l)}{\partial \lambda}\varphi_{n}(
\lambda;l){\rm d}\rho(\lambda;l)
=-\int\varphi_{n}^{2}(\lambda;l)(\frac{\partial}{\partial \lambda}{\rm d}\rho(
\lambda;l))
\nonumber \\
&=&
-\sum_{k=0}^{\infty}kt_{k}\int\lambda^{k}\varphi_{n}^{2}(\lambda;l){\rm d}\rho(\lambda;l)
+l\int\frac{\varphi_{n}^{2}(\lambda;l)}{\lambda}{\rm d}\rho(l).
\label{6.2}
\end{eqnarray}
Summing (\ref{6.2}) over $n$ from $0$ to $p-1$,
we get
\begin{equation}
L_{-1}(l)Z_{p}(l)=0.
\end{equation}
This is the lowest string equation.

Second 
from the orthonormality,
we also have
\begin{equation}
n=\int\lambda\frac{\partial\varphi_{n}(\lambda;l)}{\partial \lambda}\varphi_{n}
(\lambda;l){\rm d}
\rho(\lambda;l),
\label{se2}
\end{equation}
which  can be partially integrated as
\begin{eqnarray}
n&=&-\int\varphi_{n}(\lambda;l)\frac{\partial}{\partial \lambda}(\lambda\varphi_{n}
(\lambda;l)
{\rm d}\rho(\lambda;l))
\nonumber \\
&=&
-\int\varphi_{n}(\lambda;l)\{\varphi_{n}(\lambda;l)+\lambda\frac{\partial \varphi_{n}
(\lambda;l)}{\partial 
\lambda}+l\varphi_{n}(\lambda;l)-\sum_{k=0}^{\infty}kt_{k}\lambda^{k}\varphi_{n}(\lambda;l)\}
{\rm d}\rho(l).
\label{5.12}
\end{eqnarray}
Using again the orthonormality, we get 
\begin{equation}
2n+1+l=\sum_{k=0}^{\infty}kt_{k}\int\lambda^{k}\varphi_{n}^{2}(\lambda;l){\rm d}\rho(
\lambda;l).
\label{5.13}
\end{equation}
If we sum over (\ref{5.13})  with respect to $n$  from $0$ to $p-1$, 
we obtain
\begin{equation}
L_{0}(l)Z_{p}=0.
\end{equation}
This is the string equation for $q=0$.

\setzero

\section{Continuous Toda and  Discrete Toda Theories}

Operators $A(l)^{k}$ are not  symmetric, but one can  rewrite it in a 
symmetric form:
\begin{equation}
{\cal A}(l)^{k}=
T(l)
A(l)^{k}
T(l)^{-1}
\label{A}
\end{equation}
where $T(l)$ is
\begin{eqnarray}
\left(
\begin{array}{cccc}
e^{\phi_{0}(l)/2}& & & \bzr\\
  &e^{\phi_{1}(l)/2}& & \\
 & &\cdots& \\
\bzl& & & e^{\phi_{N-1}(l)/2}
\end{array}
 \right).
\end{eqnarray}
In fact, ${\cal A}(l)$ is a matrix representation of (\ref{l1}):
\begin{eqnarray}
\left(
\begin{array}{ccccc}
b_{0}& a_{1}& & & \bzr \\
  a_{1}&b_{1}& a_{2}& & \\
 & \cdots&\cdots& & \\
 & &\cdots&\cdots& \\
 & & a_{N-2}&b_{N-2}&a_{N-1}\\
\bzl& & & a_{N-1}& b_{N-1}
\label{ma}
\end{array}
 \right).
\end{eqnarray}
{}From the operator ${\cal A}(l)^{\infty}$ the continuous  Toda hierarchy is obtained.
 \cite{m}
The Toda vertex operator  is defined by:\cite{a}
\begin{equation}
\tilde{X}(\lambda;l)
=\exp(-\sum t_{i}\lambda^{i}) \exp(2\sum \lambda^{-i}\frac{1}{i}
\frac{\partial }{\partial t_{i}}).
\end{equation}
Using this operator we can relate 
  $Z_{p}(l)$ and $Z_{p-1}(l)$  as
\begin{equation}
Z_{p}(l)=\tilde{\cal{X}}(l)Z_{p-1}(l),
\end{equation}
where 
\begin{equation}
\tilde{\cal{X}}(l)=
\int\lambda^{2p-2-l}
\tilde{X}(l){\rm d}\lambda.
\end{equation}
On the other hand, we can construct  relation between 
$Z_{p}(l)$ and $Z_{p}(l+k)$ as follows.
First  we write  $Z_{p}(l)$ in the  bra-ket expression,
\begin{equation}
Z_{p}(l)=\prod_{i=0}^{p-1}|\Phi_{i}(\lambda;l)\rangle\langle \Phi_{i}(\lambda;l)|A(l)^{l}.
\end{equation}
We next define  operator by
\begin{equation}
\tilde{{\cal T}}^{(k)}(l)=R^{(k)}(l)\bigotimes (L^{(k)}(l))^{-1}
=(L^{(k)}(l))^{-1}\bigotimes R^{(k)}(l).
\end{equation}
 Since $w(\lambda;l+k)=\lambda^{k}w(\lambda;l)$  the relation between $Z_{p}(l)$ and $Z_{p}(l+k)$ can be expressed as
\begin{equation}
Z_{p}(l+k)=\tilde{{\cal T}}^{(k)}(l)Z_{p}(l)
\end{equation}
\setzero
\section{Concluding Remarks}

In the present letter we have put forward  to a point of view that 
matrix models are related to  to the discrete Toda theory

By introducing  a discrete time variable to the theory of 
orthogonal polynominals,
we have  found that the whole theory corresponds to 
the discrete Toda theory.
We have also shown that the $\tau$ functions  of the orthogonal polynominals 
are  equal to the partition functions of the matrix model.
This matrix model is  obtained  in the limit of the Kontsevich
 integral and in the case $l=0$ becomes  the normal one-matrix model.
To summarize, the theory of the matrix model can be described by using 
the discrete Toda theory.

\end{document}